\documentclass[twocolumn,british,english,aps,prb,showpacs,superscriptaddress,groupedaddress]{revtex4}  
\usepackage[LGR,T1]{fontenc}
\usepackage[latin9]{inputenc}
\usepackage{textcomp}
\usepackage{amstext}
\usepackage{amssymb}
\usepackage{graphicx}
\usepackage[letterpaper, margin=1.4cm]{geometry}
\makeatletter

\DeclareRobustCommand{\greektext}{%
  \fontencoding{LGR}\selectfont\def\encodingdefault{LGR}}
\DeclareRobustCommand{\textgreek}[1]{\leavevmode{\greektext #1}}
\DeclareFontEncoding{LGR}{}{}
\DeclareTextSymbol{\~}{LGR}{126}

\@ifundefined{textcolor}{}
{%
 \definecolor{BLACK}{gray}{0}
 \definecolor{WHITE}{gray}{1}
 \definecolor{RED}{rgb}{1,0,0}
 \definecolor{GREEN}{rgb}{0,1,0}
 \definecolor{BLUE}{rgb}{0,0,1}
 \definecolor{CYAN}{cmyk}{1,0,0,0}
 \definecolor{MAGENTA}{cmyk}{0,1,0,0}
 \definecolor{YELLOW}{cmyk}{0,0,1,0}
}

\def\@pacs@name{}%
\makeatother
\usepackage{babel}

\begin{document}
\title{{\LARGE{}Quantum Photonic Interconnect}}

\author{Jianwei Wang$^{1}$, Damien Bonneau$^{1}$, Matteo
Villa$^{1,2}$, Joshua W. Silverstone$^{1}$, Raffaele Santagati$^{1}$,
Shigehito Miki$^{3}$, Taro Yamashita$^{3}$, Mikio Fujiwara$^{4}$,
Masahide Sasaki$^{4}$, Hirotaka Terai$^{3}$, Michael G. Tanner$^{5}$,
Chandra M. Natarajan$^{5}$, Robert H. Hadfield$^{5}$, Jeremy L.
O\textquoteright Brien$^{1}$, and Mark G. Thompson\email{Email: Mark.Thompson@bristol.ac.uk}$^{\dagger,\,}$}

\affiliation{Centre for Quantum Photonics, H. H. Wills Physics Laboratory \& Department of Electrical and Electronic Engineering, University
of Bristol, Merchant Venturers Building, Woodland Road, Bristol BS8
1UB, United Kingdom.\\ 
$^\textrm{\textit{2 }}$Istituto di Fotonica e Nanotecnologie (IFN), Dipart di Fisica-Politecnico
di Milano, Piazza Leonardo da Vinci 32, 20133 Milano\textendash Italia.\\
$^\textrm{\textit{3 }}$National Institute of Information and Communications Technology
(NICT), 588-2 Iwaoka, Kobe 651-2492, Japan.\\ 
$^\textrm{\textit{4 }}$National Institute of Information and Communications Technology
(NICT), 4-2-1 Nukui-Kitamachi, Koganei, Tokyo 184-8795, Japan. \\
$^\textrm{\textit{5 }}$School of Engineering, University of Glasgow, G12 8QQ, United
Kingdom.\\
\vspace{0.04cm}
$^\dagger${Please correspondence to: mark.thompson@bristol.ac.uk}
}

\begin{abstract}
{\normalsize{}Integrated photonics has enabled much progress towards quantum technologies. Many applications, including quantum communication, sensing, and distributed and cloud quantum computing, will require coherent photonic interconnection between separate chip-based sub-systems. Large-scale quantum computing systems and architectures may ultimately require quantum interconnects to enable scaling beyond the limits of a single wafer and towards "multi-chip" systems. However, coherently interconnecting separate chips is challenging due to the fragility of these quantum states and the demanding challenges of transmitting photons in at least two media within a single coherent system. Distribution and manipulation of qubit entanglement between multiple devices is one of the most stringent requirements of the interconnected system. Here, we report a quantum photonic interconnect demonstrating high-fidelity entanglement distribution and manipulation between two separate chips, implemented using state-of-the-art silicon photonics. Path-entangled states are generated and manipulated on-chip, and distributed between the chips by interconverting between path-encoding and polarisation-encoding. We use integrated state analysers to confirm a Bell-type violation of $S$=2.638\textpm0.039 between two chips. With improvements in loss, this quantum interconnect will provide new levels of flexible systems and architectures for quantum technologies.}
\end{abstract}
\maketitle


\begin{figure*}[t]
\begin{centering}
\includegraphics[scale=0.74]{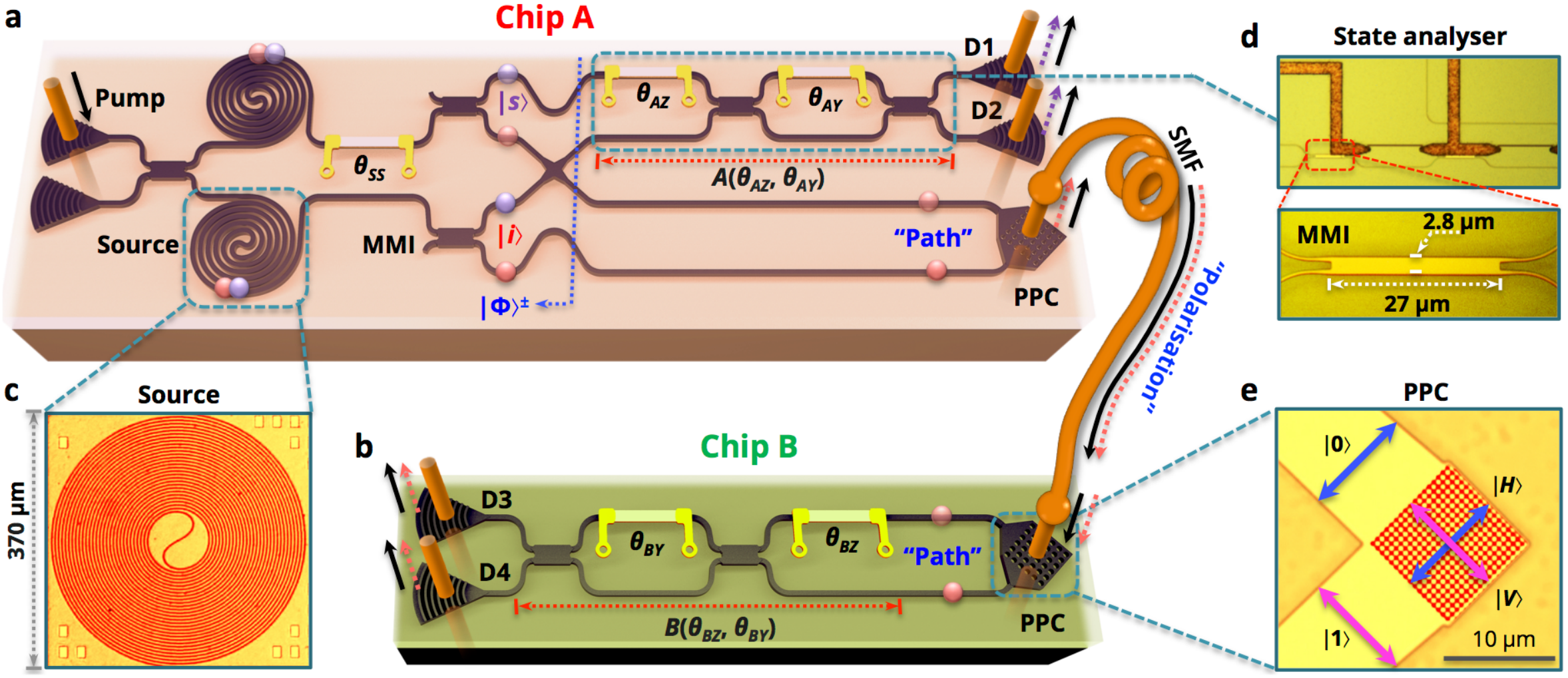}
\par\end{centering}
\centering{}\protect\caption{\textbf{Quantum photonic interconnect and entanglement distribution between two integrated Si photonic chips}. 
\textbf{a}, Chip-A comprises path-entangled states generation, arbitrary projective measurement $A(\theta_{AZ},\theta_{AY})$, and path/polarisation conversion (PPC). \textbf{b}, Chip-B includes a projective measurement $B(\theta_{BZ},\theta_{BY})$ and PPC. On the chip-A,  signal-idler photon-pairs are created in the spiralled waveguide single-photon source. Bell states $|\Phi\rangle^{\pm}$ are produced when $\theta_{SS}$ is controlled to be $\pi/2$ or $\pi$. Idler qubit initially encoded in path are coherently coupled to polarisation-encoding and transmitted through a 10m single-mode optical fibre (SMF), and reversely converted back to path-encoding on the chip-B. Signal qubit is analysed using $A(\theta_{AZ},\theta_{AY})$ on chip-A and idler qubit is analysed using $B(\theta_{BZ},\theta_{BY})$ on chip-B. The 2D grating coupler, behaving as the path/polarisation converter (PPC), is used to coherently interconvert photonic qubit between polarisation and path-encoding. Optical microscopy images of, \textbf{c}, the photon-pair source, \textbf{d}, the arbitrary state analyser (Inset shows the MMI splitter), and \textbf{e}, the 2D grating coupler PPC structure. }
\end{figure*}
\vspace{0.03cm}

  
Further progress towards quantum communication \cite{Kimble, Gisinqc-np-2007}, sensing \cite{MetrologyRev} and computing \cite{Ladd2010Quantum,Jeremy} will greatly benefit from a "quantum photonic interconnect": an inter-$/$intra-chip link---e.g. in optical-fibre or free-space---capable of {\em coherently} distributing quantum information and entanglement between on-chip sub-systems within a single complete quantum system. The significance of quantum interconnect was first highlighted by {\em Kimble}\cite{Kimble}, and here we study a chip-based interconnect solution, which will be essential in many future applications and provide substantial architectural flexibility. Secure quantum key distribution and quantum communication \cite{Pan100teleportation,Xiaosong, Ursin:2004jb}, and distributed ad cloud quantum computing \cite{St-Science-335-303, Fisher, DistributedQC} require interconnected on-chip subsystems for practical implementations. Precise quantum sensing will gain flexibility and versatility from on-chip generation and detection of entanglement, with the interaction with sample performed in a different media or location, e.g., chip, optical fiber and free space\cite{Matt-NP-3-346,protein, Bowen}. Quantum computing will benefit from quantum interconnects through architectural simplifications \cite{Benjamin, Gabriel, Terry}; easier integration of materials and platforms optimised for source \cite{GaAs, cl-oe-17-16561}, circuit \cite{Po-Science-320-646, Shadbolt-np-2011, SilverstoneSi, SilverstonBell, Metcalf, Paolo.Ncom}, detector \cite{Pernice:2012p11448, Drik.Detector} and many others \cite{Toshiba.review, Dirk.Filter} performance; and the inclusion of off-chip optical delays or memories. Ultimately, large-scale integrated quantum systems may even exceed the area of a single wafer or require interconnects for architectural reasons.

A  quantum photonic interconnect must maintain coherent transmission of the qubit state $\alpha\left|0\right\rangle+\beta\left|1\right\rangle$ between subsystems; a significant difference compared to the classical optical interconnect in which transmitted state are either 0 or 1, and in which the relative phase is not maintained\cite{Vlasov}. The quantum interconnect must also be capable of coherently interconverting between the preferred encodings in the platforms and media through which it connects\cite{NielsenChuang, Kimble}. Perhaps, the most demanding requirement for an interconnect is the preservation of high-fidelity entanglement throughout any manipulation, interconversion and transmission processes within the full system. Distributing entanglement\cite{RevModPhys.Entanglement} between integrated chips is a key requirement and a major technical challenge due to the highly fragile nature of entanglement and the potential for decoherence of quantum states transmitted between different chips. 
Path-encoding\cite{Po-Science-320-646, Shadbolt-np-2011, SilverstoneSi, SilverstonBell, Metcalf}---a photon across two waveguides---is a natural choice of robust on-chip encoding for quantum information processing, however, polarisation\cite{Pan100teleportation,Xiaosong, Ursin:2004jb, St-Science-335-303, Fisher}, spatial-mode\cite{Wang:dma, DAmbrosio:2012bk}, or time-bin\cite{Time.Inagaki} encoding is typically more suitable in fibre and free space for quantum information transmission and distribution. Already there have been demonstrations of important features of quantum interconnect components, including on-chip entanglement generation and manipulation\cite{Po-Science-320-646, Shadbolt-np-2011, SilverstoneSi, SilverstonBell, Metcalf, Paolo.Ncom, Matsuda, Olislager:13}, photon detection\cite{Pernice:2012p11448, Drik.Detector}, interfacing of light's different degrees of freedom\cite{2Dcoupler, DaiRev, IOAM}, and multi-chip links\cite{Toshiba.review, Dirk.Filter}. However, to date there has been no demonstration of a complete quantum photonic interconnect system capable of coherently distributing and manipulating qubit entanglement across two or more integrated quantum circuits.

Here, we demonstrate a high-fidelity quantum photonic interconnect. Telecom-band entangled photons are generated, manipulated and distributed between two integrated silicon photonic chips linked by an optical-fibre. These chips were fabricated using state-of-the-art technologies from silicon photonics to enable and monolithically integrate all of the capabilities required for the quantum interconnect. Path-entangled Bell states are generated and manipulated on-chip. These entangled-states are distributed across two chips, by transmitting one qubit from one chip to the other via the fibre. Coherence is preserved by interconverting between path-encoding on chips and polarisation-encoding in fibre using a two-dimensional grating coupler \cite{2Dcoupler}. Each qubit is analysed in the respective chips using thermal phase shifters to form integrated state analysers. We show a high-fidelity interconversion from path-encoding on one chip to polarisation-encoding within the fibre, and back to path-encoding on the second chip. We implement a rigorous test of entanglement---confirming a strong Bell-type inequality violation of 16.4$\sigma$ or 15.3$\sigma$---and demonstrate the quantum interconnect. Together with further improvement in loss, this approach will provide new quantum technologies and applications that rely on or benefit from quantum photonic interconnects. 

Figure 1 shows the schematics of a chip-to-chip quantum photonic interconnect, generating path-entangled states on chip-A and coherently distributing one entangled qubit to chip-B, via a 10m single mode optical fibre link. Chip-A and chip-B respectively have an effective footprint of 1.2\texttimes{}0.5 mm$^{2}$ and 0.3\texttimes{}0.05 mm$^{2}$. On chip-A, a signal ($\lambda_{s}$$\sim$1550.7 nm) and idler ($\lambda_{i}$$\sim$1560.3 nm) photon pair is generated via the elastic scattering of two photons from a bright continuous-wave pump field ($\lambda_{p}$$\sim$1555.5 nm) inside 2cm spiralled waveguide sources (Fig.1c), by using the spontaneous four-wave-mixing (SFWM) nonlinear effect\cite{cl-oe-17-16561}. The pump is split across two sources using a multimode interference (MMI) beam splitter with a near 50/50 splitting ratio\cite{Damien.Si} (Fig.1d). The photon pairs are produced in either the top or bottom waveguides, yielding a photon-number entangled state\cite{SilverstoneSi} as $(\left|1_{s}1_{i}\right\rangle \left|0_{s}0_{i}\right\rangle -e{}^{2i\theta_{SS}}\left|0_{s}0_{i}\right\rangle \left|1_{s}1_{i}\right\rangle )/\sqrt{2}$, where $\theta_{SS}$ is a thermally-controlled phase after the sources. These photons are probabilistically separated by two demultiplexing MMIs and post-selected by two off-chip filters, producing the maximally path-entangled Bell states $\left|\Phi\right\rangle ^{\pm}=(\left|0\right\rangle _{s}\left|0\right\rangle _{i}\pm\left|1\right\rangle _{s}\left|1\right\rangle _{i})/\sqrt{2}$, with a 25\% success probability, when $\theta_{SS}$ equals to $(n+1/2)\pi$ or $n\pi$ for an integer $n$. Subscript \textit{s} and \textit{i} represent the logical states of signal and idler qubits (more details see Supplementary Information). Then, we use an on-chip path/polarisation converter (PPC) to coherently interconvert the idler qubit between its path and polarisation-encoding. On chip-A, path-encoded qubit is converted to polarisation-encoded before transmitting across the fibre. Chip-B reverses this process, converting the polarisation-encoded qubit back to on-chip path-encoded qubit, by using a PPC. This PPC enables entanglement preservation throughout the chip and fibre platforms. Signal and idler qubits are manipulated and measured independently on two chips using arbitrary single qubit measurement stages $A(\theta_{AZ},\theta_{AY})$ and $B(\theta_{BZ},\theta_{BY})$, which physically consists of an on-chip Mach-Zehnder interferometer with an additional thermal phase shifter (Fig.1d).

We first discuss the coherent interconversion of path and polarisation-encoding by using the PPC. In silicon quantum photonics, transverse-electric (TE) mode is usually in use, owing to its stronger waveguide confinement and consequently enhanced SFWM effect\cite{Matsuda}. Thus, produced photons must be guided in TE-modes, which is easily achieved by injecting pump light in this mode using a 1D TE-grating coupler. Our PPC is implemented using a 2D grating coupler (Fig.1e), where TE-polarised light coming from two nearly orthogonal waveguides is combined into two orthogonal polarised components of light\cite{2Dcoupler, Olislager:13}. In this way, the polarisation states of photons received by the fibre is determined by the two-waveguide on-chip states, and vice versa. This provides a coherent interconversion between path-encoding and polarisation-encoding. Details are provided in the Methods and SI. To confirm the PPC coherent mapping, we prepared arbitrary bright-light polarisation states using bulk optical components and coupled them into the on-chip receiver (Fig.2a). The 2D grating coupler converted the polarisation states into path-encoded states, which were then analysed on-chip by implementing a full state tomography\cite{NielsenChuang}. We prepared a set of six polarisation states $\rho_{pol}$, and measured the corresponding on-chip path states $\rho_{path}$; these states are shown as Bloch (or Poincare) vectors in Figures 2b and 2c, respectively. The distance between the states can be described by the state fidelity, which is defined as $F_{state}=(Tr[\sqrt{\sqrt{\rho_{pol}}\cdot\rho_{path}\cdot\sqrt{\rho_{pol}}}])^{2}$. The mean fidelity of the six measured states is 98.82\textpm 0.73\%. An example of a reconstructed density matrix for the path-encoded state $|+\rangle$ is shown in Fig. 2d (full data are provided in Fig.S5). Then, we fully quantified the PPC process using a quantum process tomography\cite{NielsenChuang}. This is mathematically described by a process matrix $\chi$, defined by $\rho_{path}=\sum_{mn}(E_{m}\rho_{pol}E_{n}^{\dagger}\chi_{mn})$, where $E_{i}$ are the Identity matrix \textit{I} and Pauli matrices \textit{X}, \textit{Y}, and \textit{Z}, respectively. By subjecting the $\rho_{pol}$ states into the PPC and measuring the $\rho_{path}$ states, we determined the process matrix $\chi$ of the PPC, shown in Fig.2e. We find a high process fidelity of 98.24\textpm 0.82\%, defined as $F_{process}=Tr[\chi_{ideal}\cdot\chi]$, where $\chi_{ideal}$ is the ideal process matrix with $\chi_{ideal}[I,I]=1$. \textit{X}, \textit{Y}, and \textit{Z} amplitudes of the measured matrix $\chi$ represent the probability of a bit-flip or phase-flip error on the PPC interconversion. 

\begin{figure}[t]
\begin{centering}
\includegraphics[scale=0.4]{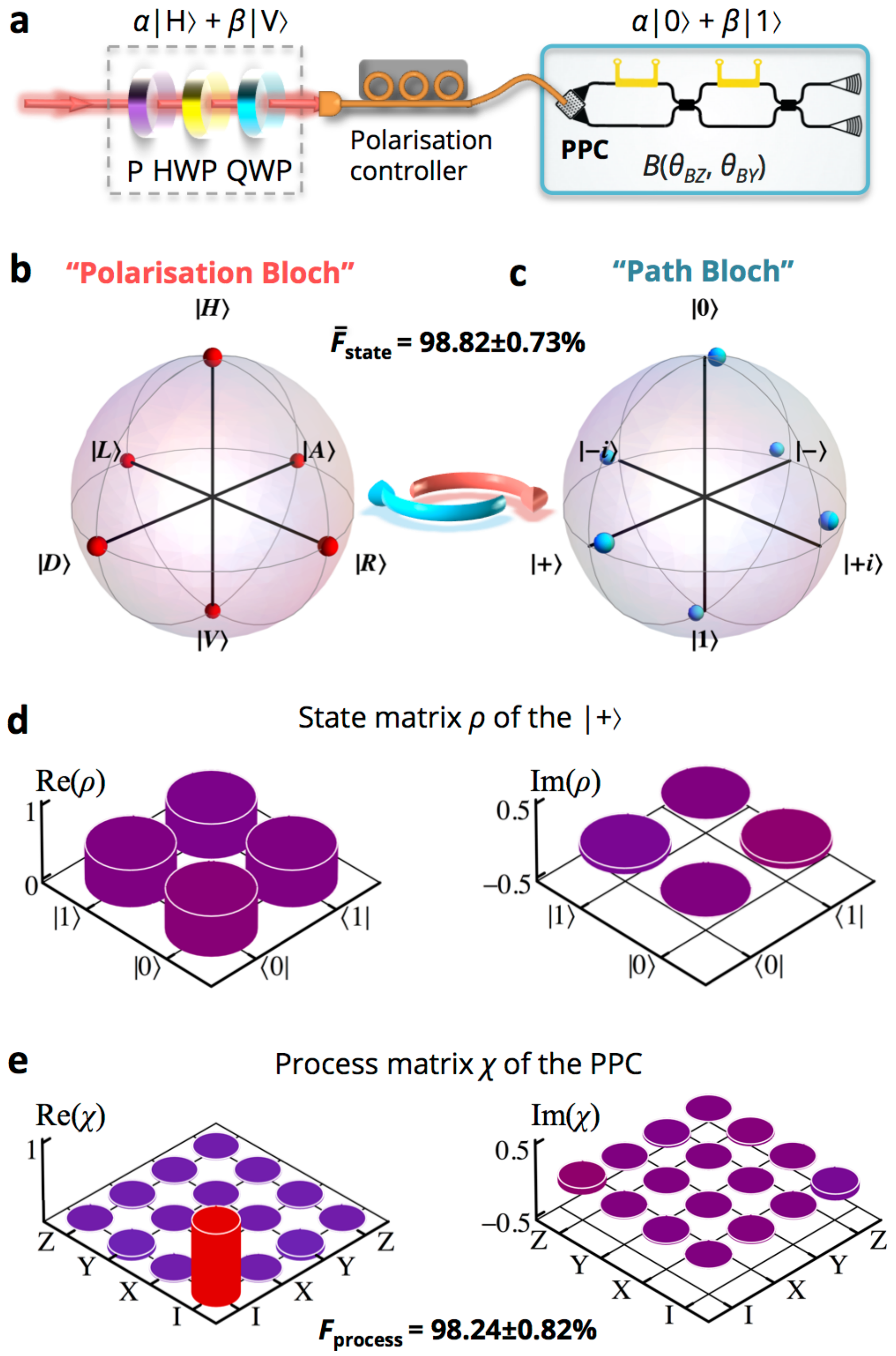}
\par\end{centering}
\centering{}\protect\caption{\textbf{Interconversion between polarisation-encoding and path-encoding}.
\textbf{a}, Initial arbitrary polarisation-encoded states $\alpha|H\rangle+\beta|V\rangle$ ($|H\rangle$ and $|V\rangle$ are two orthogonally polarised-states) were prepared by using a set of polariser (P), half-wave plate (HWP) and quarter-wave plate (QWP). A fibre polarisation controller was used to compensate polarisation rotation in the fibre. The PPC converted polarisation states into path-encoded states $\alpha|0\rangle+\beta|1\rangle$, where $|0\rangle$ and $|1\rangle$ denote path states in two waveguides. The path-encoded states were analysed using $B(\theta_{BZ},\theta_{BY})$ to implement state tomography. \textbf{b} and \textbf{c}, the Bloch sphere representation of ideal polarisation-encoded states $\{|H\rangle,|V\rangle,|D\rangle,|A\rangle,|R\rangle,|L\rangle\}$ in bulk optics, and measured path-encoded states $\{|0\rangle,|1\rangle,|+\rangle,|-\rangle,|+i\rangle,|-i\rangle\}$ on chip. Indicated fidelity represents the mean over all six states. \textbf{d}, Reconstructed density matrix of the $|+\rangle$ path-encoded state corresponding to the $|D\rangle$ polarisation-encoded state. \textbf{e}, Reconstructed process matrix $\chi$ of the PPC using the quantum process tomography. }
\end{figure}

A $\sim$50 mW CW pump was injected into on-chip sources to create photon pairs. Photons were detected using two superconducting nanowire single-photon detectors (SNSPDs) with $\sim$50\% efficiencies and $\sim$800 Hz dark counts. Coincidences were recorded using a time interval analyser. After the chip-A, a mean rate of 500\textminus 800 Hz photon pairs was observed, while after the two chips we obtained 8\textminus 12 Hz mean coincidences. The pump light propagates collinearly with single photons, and this allows a closed feedback loop to track photons and monitor state stability throughout the chip-to-chip quantum interconnect system. Details are provided in the Methods and SI. 

We next configured chip-A to produce entangled states. Signal and idler photons were respectively collected at ports D1 and D2 of the chip-A, and routed to SNSPDs. Continually scanning $\theta_{SS}$, we observed the $\lambda$-classical interference and $\lambda$/2-quantum interference fringes with a visibility ($V=1-N_{min}/N_{max}$) of 99.99\textpm 0.01\% and 99.36\textpm 0.17\%, respectively (Fig. 3a). The high visibility of this double-frequency fringe is a signature of high-quality photon-number entanglement produced inside the chip-A\cite{Shadbolt-np-2011, Matt-NP-3-346}. The high visibilities arise from well-balanced MMI splitters\cite{Damien.Si} and a good spectral overlap between two photon-pair sources\cite{SilverstoneSi, SilverstonBell}.
Then, the photon-number entangled state evolves into the path-entangled Bell states, $|\Phi\rangle^{+}$ or $|\Phi\rangle^{-}$, by setting $\theta_{SS}$ to $\pi/2$ or $\pi$. The entangled-qubits were separated and coherently distributed across chip-A and chip-B using the PPC interconversion. We measured correlation fringes across the two chips, by collecting signal photons at port D1 on chip-A and idler photons at port D3 on chip-B, and simultaneously operating $A(\theta_{AZ}, \theta_{AY})$ and $B(\theta_{BZ}, \theta_{BY})$ on two chips. Figures 3b and 3c show the entanglement correlation fringes for the Bell states $|\Phi\rangle^{+}$ and $|\Phi\rangle^{-}$ as a rotation of $\theta_{BY}$ on chip-B, with $\theta_{AY}$ on chip-A set at \{$0$, $\pi/2$, $\pi$, $3\pi/2$\}. These results are in good agreement with their theoretical predictions of $cos^{2}[(\theta_{AY}-\theta_{BY})/2]$ and $cos^{2}[(\theta_{AY}+\theta_{BY})/2]$\cite{Kwiat.source}. The fringes exhibit a mean visibility of 97.63\textpm 0.39\% and 96.85\textpm 0.51\%, respectively, which is far beyond the critical visibility required of $1/\sqrt{2}$ to violate the Bell inequality\cite{PhysRevLett.64.2495}. We now have shown that a very high-quality entanglement is produced on the chip-A, and distributed over the fibre to the chip-B, coherently interconnecting the two chips. 

To more strictly verify the existence of entanglement across the two chips, we directly measured the Bell-CHSH (Clauser-Horne-Shimony-Holt) inequality\cite{Holt69}, defined as: 
\begin{equation}
S=\Vert\langle A_{1}, B_{1}\rangle+\langle A_{1}, B_{2}\rangle+\langle A_{2}, B_{1}\rangle-\langle A_{2}, B_{2}\rangle\Vert\leq2
\end{equation}
where $A_{i}$ and $B_{i}$ briefly denote the projectors $A(0,\theta_{AY})$ and $B(0,\theta_{BY})$ on the two chips. Correlation coefficients $\langle A_{i}, B_{i}\rangle$ were measured, when $\theta_{AY}$ on chip-A was set to $\{0, \pi/2\}$ and $\theta_{BY}$ on chip-B was set to $\{\pi/4, 3\pi/4\}$. Full data of $\langle A_{i}, B_{i}\rangle$ is provided in Fig.S6. Using equation (1), we obtained the directly measured $S_{CHSH}$ parameters of 2.638\textpm 0.039 and 2.628\textpm 0.041 for the two Bell states $|\Phi\rangle^{+}$ and $|\Phi\rangle^{-}$, respectively. These $S_{CHSH}$ violate the Bell-CHSH inequality by 16.4 and 15.3 standard deviations, strongly confirming that the two photons after distributed across chip-A and chip-B are highly entangled, and therefore verifying a high-quality quantum photonic interconnect between two chips. In addition, we estimate the maximally achievable $S_{fringe}$ parameters of 2.761\textpm 0.011 and 2.739\textpm 0.015 for the $|\Phi\rangle^{+}$ and $|\Phi\rangle^{-}$ states, from the mean visibility of the correlation fringes (shown in Fig.3) according to $S_{fringe}=2\sqrt{2}V$\cite{PhysRevLett.64.2495}. Figure 4 illustrates a good agreement of the $S_{CHSH}$ and $S_{fringe}$ parameters. 

\begin{figure}
\begin{centering}
~\includegraphics[scale=0.85]{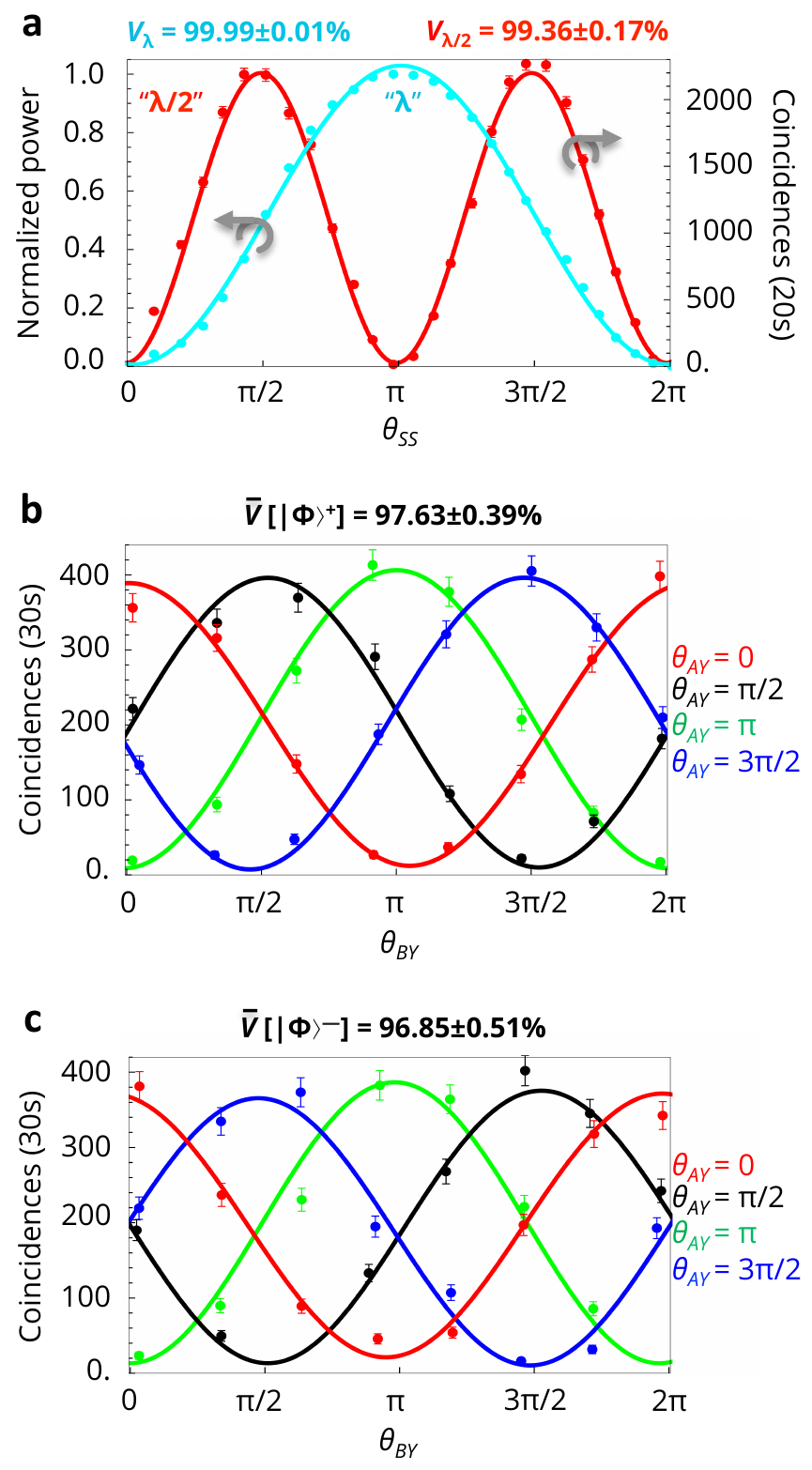}
\par\end{centering}
\centering{}\protect\caption{\textbf{Entanglement fringes}. 
\textbf{a}, $\lambda$-classical interference (cyan) and $\lambda$/2-quantum interference (red) fringes measured on chip-A. Bright-light was measured (normalised) at port D1, and coincidences were collected (accumulated 20s) between ports D1 and D2. $\theta_{SS}$ was rotated to produce fringes, as $A(\theta_{AZ},\theta_{AY})$ was set as a Hadamard gate. Photons are bunched or anti-bunched when $\theta_{SS}$ is $n\pi$ or $(n+1/2)\pi$. \textbf{b} and \textbf{c}, Entanglement correlation fringes for the Bell states $|\Phi\rangle^{+}$ and $|\Phi\rangle^{-}$ after distributed across the two chips. Coincidences were collected (accumulated 30s) between ports D1 and D3. $\theta_{BY}$ on chip-B was continually rotated ($\theta_{BZ}=0$) to obtain the fringes, as $A(\theta_{AZ},\theta_{AY})$ on chip-A was projected onto $\{|1\rangle,|0\rangle,|-\rangle,|+\rangle\}$ basis by setting $\theta_{AY}$ to $\{0, \pi/2,\pi, 3\pi/2\}$ and $\theta_{AZ}$ to 0. The indicated visibility represents the mean over all four fringes. Error bars are given by Poissonian statistics. Accidental coincidences are subtracted.}
\end{figure}

\begin{figure}[t]
\begin{centering}
\includegraphics[scale=0.35]{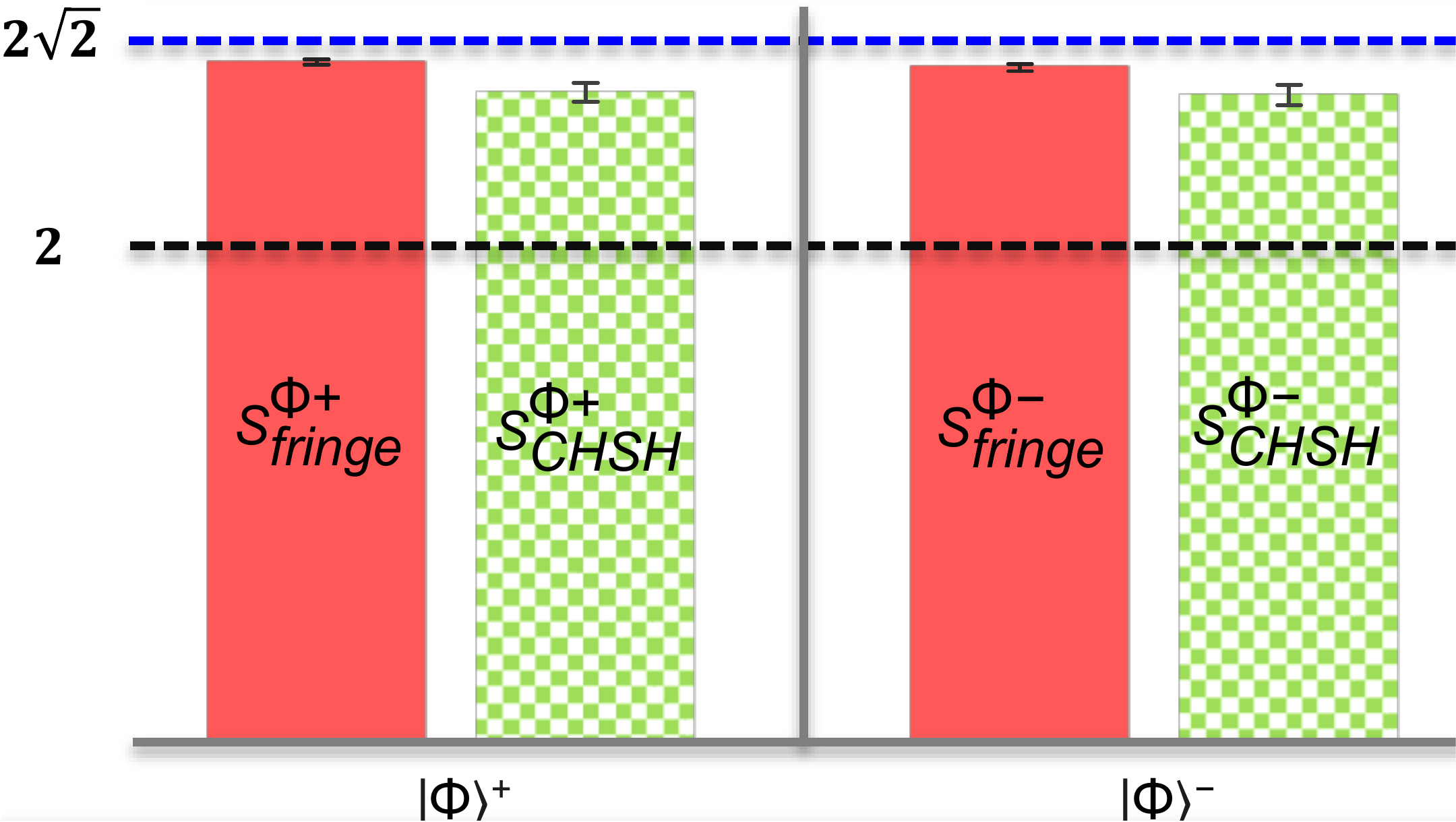}
\par\end{centering}
\centering{}\protect\caption{\textbf{Verification of chip-to-chip entanglement distribution and quantum photonic interconnect.} 
The two Bell states were distributed across two Si chips. 
The \textit{S} parameters were obtained using two approaches. Green dotted columns are the directly measured $S_{CHSH}$ using Eq. (1). Pink columns are the maximal achievable $S_{fringe}$, estimated from the mean visibility of correlation fringes in Fig.3. The $S_{CHSH}$ and $S_{fringe}$ parameters are in good agreement. 
Black and blue dashed lines denote the classical and quantum boundary. These results confirm the high levels of entanglement after distributed across chips, and the high fidelity of quantum photonic interconnect. Coincidences of each measurement were accumulated for 60s and accidental coincidences are subtracted. Error bars ($\pm1$ s.d.) are given by Poissonian statistics.}
\end{figure}

We demonstrate high-fidelity entanglement generation, manipulation, interconversion, distribution and measurement across two separate integrated photonic circuits, successfully demonstrating the first chip-to-chip quantum photonic interconnect. Path--polarization interconversion preserves quantum coherence across the full interconnected chip-fiber-chip system, demonstrating beyond single-chip implementation of a quantum photonic experiment. The efficiency of this interconversion process can be further improved by engineering the geometry of grating coupler or utilising other polarisation control techniques \cite{DaiRev, Ding}. Other interconversion approaches, e.g., path--orbital angular momentum \cite{Path-OAM, TOAM} or path--time bins \cite{Cudos} may further enrich this quantum photonic interconnectivity. The use of silicon---allowing large-scale integration\cite{Sun} and compatibility with microelectronics and telecommunications infrastructure\cite{Bogaerts, Vlasov}, benefiting from the classical optical interconnect technology on silicon \cite{Vlasov, DaiRev}, and also offering ability to monolithically integrate photon source\cite{cl-oe-17-16561, SilverstoneSi}, circuit\cite{Damien.Si, SilverstonBell} and detector\cite{Pernice:2012p11448,Drik.Detector}---would position this quantum photonic interconnect technology at the heart of building practical, robust and scalable silicon-based quantum hardware for future and networks. This work opens the door to multi-chip integrated quantum photonic technologies, which would be capable of processing quantum information on extremely small and stable chips and also capable of robustly distributing and transmitting quantum information between chips. 

\section*{{\large{}Methods}}
\textbf{\small{}Devices design and fabrication}{\small{}. The devices were fabricated on the standard silicon-on-insulator wafer with a 220 nm silicon layer and a 2 \textgreek{m}m buried silica oxide layer. MMI couplers were designed as 2.8 \textmu m\texttimes{}27 \textmu m to get a balanced splitting ratio (Fig.1d). MMIs offer a large bandwidth and a large fabrication tolerance. Spiralled waveguide sources with a 2-cm length were used to create photon-pairs. The 1D grating couplers consist of a periodic 315 nm silicon layer with a 630 nm pitch. The 2D grating couplers include 10 \textmu m\texttimes{}10 \textmu m hole arrays with a 390 nm diameter and a 605 nm pitch. Resistive heaters with a 50 \textmu m-length were designed and formed by a Ti/TiN metal layer. The devices were fabricated using the deep-UV (193 nm) lithography at LETI-ePIXfab. Silicon waveguides were 220 nm fully etched, while 1D and 2D grating couplers were 70 nm shallow etched. The devices were covered by a 1.6 \textmu m silica oxide layer. }{\small \par}

\textbf{\small{}Devices characterisations.}{\small{} Optical accesses and electric accesses were independently controlled on two chips (Fig.S1). Optical accesses were achieved using V-groove single modes fibre arrays with a 127 \textmu m pitch. Fibres were titled with an angle of 10}$\thicksim${\small{}12 degrees to guarantee both grating couplers work at the required wavelengths. Excess loss of 1D and 2D grating couplers were about -4.8 dB and -7.6 dB at peak wavelengths, respectively (Fig.S2). Extinction ratio of 1D and 2D grating couplers were measured to be larger than 20 dB and 18 dB, respectively. We
estimated losses from different contributors in the system: -6 dB from off-chip filters, -6 dB from SSNPDs, -9}$\thicksim${\small{}9.5 dB from 1D grating couplers, -15}$\thicksim${\small{}15.5 dB from 2D grating couplers, -6 dB from demultiplexing MMIs, and -8}$\thicksim${\small{}9 dB from MMIs excess loss and propagation loss in waveguides. Totally, signal and idler photons respectively experienced -36}$\thicksim${\small{}38 dB and -18}$\thicksim${\small{}19 dB attention. The fibre channel interconnected two chips has -15}$\thicksim${\small{}16 dB attention mainly from the 2D grating couplers, which can be further improved by engineering the geometry of coupler and waveguide. 

All thermal-driven phase shifters were controlled using home-made electric controllers. Wire bounding technology was used to contact heaters\textquoteright{} transmission lines. Optical power was recorded as a function of electric power added on heaters. The optical---electric power contour was fitted and used to construct the mapping between the required states and electric power. Fig.S3 shows calibration results of chip-A\textquoteright s and chip-B\textquoteright s state analysers. To avoid the influence of temperature variation, both chips were mounted on temperature stabilised stages. Fibre alignment was automatically recoupled using piezo-electronic stacks. Fig.S4 shows the stability of the chip-to-chip system, which were maintained constant more than 30 mins. This indicates path-encoded states on the two chips are very stable and polarisation-encoded states in the fibre channel are also well-stabilised.}{\footnotesize{} }{\footnotesize \par}

\section*{{\large{}Acknowledgements}}
{\small{}We thank Marshall, G. D. and Murray, W. A. for experiment assistance, 
and Erven, C., Shadbolt, P. J., Matthews, J. C. F., Turner, P., Kling, L., Dai, D., and Cai, X. for useful discussions. 
This work was supported by the Engineering and Physical Science Research Council (EPSRC, UK), the European Research Council, the
Bristol Centre for Nanoscience and Quantum Information, the European FP7 project BBOI, UK Quantum Communications Hub project, and the ImPACT Program of the Cabinet Office Japan. J.L.O'B. acknowledges a Royal Society Wolfson Merit Award and a Royal Academy of Engineering Chair in Emerging Technologies. M.G.Th. acknowledges support from an Engineering and Physical Sciences Research Council (EPSRC) Early Career Fellowship. 

}{\small \par}

\section*{{\large{}Contributions}}
{\small{}J.W., D.B. and M.G.Th. conceived the idea and designed the devices. J.W. D.B., M.V. J.W.S, and R.S. carried out the experiments.
J.W., D.B., M.V., J.W.S, R.S., J.L.O'B, and M.G.Th. analysed the experimental data. S.M., T.Y., M.F., M.S., H.T., M.G.T, C.M.N., and R.H.H. built
the single-photon detector system. All authors contributed to the manuscript.}{\small \par}

\section*{{\large{}References}}

\newpage

\onecolumngrid
\newpage
%
%

\twocolumngrid \appendix

\section{{Quantum state evolution}}

A pair of signal ($\lambda_{s}$) and idler ($\lambda_{i}$) photons
are created in the spiralled waveguide source by annihilating two
pump photons ($\lambda_{p}$), based on the $\chi(3)$ SFWM nonlinear
optical effect {[}1, 2{]}, whose Hamiltonian is approximately described
as: 

\begin{equation}
\hat{H}\text{\ensuremath{\propto}}a_{s}^{\text{\dag}}a_{p}^{\text{2}}a_{i}^{\text{\dag}}+a_{s}a_{p}^{\text{\dag}2}a_{i}
\end{equation}

where $a_{s}^{\dagger}$, $a_{p}^{\dagger}$, and $a_{i}^{\dagger}$
are the creation operators, and $a_{s}$, $a_{p}$, and $a_{i}$ are
the annihilation operators, for the involved signal, pump, and idler
photons. The CW bright pump-light is equally split into two spiralled
sources to produce a pair of signal and idler photons, which are coherently
bunched either at the top or bottom waveguides as {[}3, 4{]}: 

\begin{equation}
\left|\phi\right\rangle =(\left|1_{s}1_{i}\right\rangle _{t}\left|0_{s}0_{i}\right\rangle _{b}-e^{(\theta_{i}+\theta_{s})i}\left|0_{s}0_{i}\right\rangle _{t}\left|1_{s}1_{i}\right\rangle _{b})/\sqrt{2}
\end{equation}

where 0 and 1 represent photon number at the top and bottom waveguides
with subscript \textit{t} and \textit{b}, respectively. $\theta_{i}$
and $\theta_{s}$ are phases of the signal and idler photons controlled
by the $\theta_{SS}$ phase shifter. The difference between
$\theta_{i}$ and $\theta_{s}$ is negligible and the state can be
simply rewritten as the typical NOON states format:

\begin{equation}
\left|\phi\right\rangle =(\left|2\right\rangle _{t}\left|0\right\rangle _{b}-e^{2\theta_{SS}i}\left|0\right\rangle _{t}\left|2\right\rangle _{b})/\sqrt{2}
\end{equation}

Then, signal and idler photons are split by the demultiplexing MMI
couplers. Above photon-number entanglement state evolves to: 

\begin{eqnarray*}
\left|\phi\right\rangle  & = & \{[i\left|1_{s}1_{i}\right\rangle _{tt}\left|0_{s}0_{i}\right\rangle _{tb}\left|0_{s}0_{i}\right\rangle _{bt}\left|0_{s}0_{i}\right\rangle _{bb}\\
 &  & -i\left|0_{s}0_{i}\right\rangle _{tt}\left|1_{s}1_{i}\right\rangle _{tb}\left|0_{s}0_{i}\right\rangle _{bt}\left|0_{s}0_{i}\right\rangle _{bb}\\
 &  & +\left|0_{s}1_{i}\right\rangle _{tt}\left|1_{s}0_{i}\right\rangle _{tb}\left|0_{s}0_{i}\right\rangle _{bt}\left|0_{s}0_{i}\right\rangle _{bb}\\
 &  & +\left|1_{s}0_{i}\right\rangle _{tt}\left|0_{s}1_{i}\right\rangle _{tb}\left|0_{s}0_{i}\right\rangle _{bt}\left|0_{s}0_{i}\right\rangle _{bb}]\\
 &  & -e^{2i\theta_{SS}}[i\left|0_{s}0_{i}\right\rangle _{tt}\left|0_{s}0_{i}\right\rangle _{tb}\left|1_{s}1_{i}\right\rangle _{bt}\left|0_{s}0_{i}\right\rangle _{bb}\\
 &  & -i\left|0_{s}0_{i}\right\rangle _{tt}\left|0_{s}0_{i}\right\rangle _{tb}\left|0_{s}0_{i}\right\rangle _{bt}\left|1_{s}1_{i}\right\rangle _{bb}\\
 &  & +\left|0_{s}0_{i}\right\rangle _{tt}\left|0_{s}0_{i}\right\rangle _{tb}\left|0_{s}1_{i}\right\rangle _{bt}\left|1_{s}0_{i}\right\rangle _{bb}\\
 &  & +\left|0_{s}0_{i}\right\rangle _{tt}\left|0_{s}0_{i}\right\rangle _{tb}\left|1_{s}0_{i}\right\rangle _{bt}\left|0_{s}1_{i}\right\rangle _{bb}]\}/2\sqrt{2}
\end{eqnarray*}

And after the crossing, the state can be rewritten as: 

\begin{eqnarray*}
\left|\phi\right\rangle  & = & \{[i\left|1_{s}1_{i}\right\rangle _{tt}\left|0_{s}0_{i}\right\rangle _{tb}\left|0_{s}0_{i}\right\rangle _{bt}\left|0_{s}0_{i}\right\rangle _{bb}\\
 &  & -i\left|0_{s}0_{i}\right\rangle _{tt}\left|0_{s}0_{i}\right\rangle _{tb}\left|1_{s}1_{i}\right\rangle _{bt}\left|0_{s}0_{i}\right\rangle _{bb}\\
 &  & +\left|0_{s}1_{i}\right\rangle _{tt}\left|0_{s}0_{i}\right\rangle _{tb}\left|1_{s}0_{i}\right\rangle _{bt}\left|0_{s}0_{i}\right\rangle _{bb}\\
 &  & +\left|1_{s}0_{i}\right\rangle _{tt}\left|0_{s}0_{i}\right\rangle _{tb}\left|0_{s}1_{i}\right\rangle _{bt}\left|0_{s}0_{i}\right\rangle _{bb}]\\
 &  & -e^{2i\theta_{SS}}[i\left|0_{s}0_{i}\right\rangle _{tt}\left|1_{s}1_{i}\right\rangle _{tb}\left|0_{s}0_{i}\right\rangle _{bt}\left|0_{s}0_{i}\right\rangle _{bb}\\
 &  & -i\left|0_{s}0_{i}\right\rangle _{tt}\left|0_{s}0_{i}\right\rangle _{tb}\left|0_{s}0_{i}\right\rangle _{bt}\left|1_{s}1_{i}\right\rangle _{bb}\\
 &  & +\left|0_{s}0_{i}\right\rangle _{tt}\left|0_{s}1_{i}\right\rangle _{tb}\left|0_{s}0_{i}\right\rangle _{bt}\left|1_{s}0_{i}\right\rangle _{bb}\\
 &  & +\left|0_{s}0_{i}\right\rangle _{tt}\left|1_{s}0_{i}\right\rangle _{tb}\left|0_{s}0_{i}\right\rangle _{bt}\left|0_{s}1_{i}\right\rangle _{bb}]\}/2\sqrt{2}\\
\end{eqnarray*}

The signal and idler photons are respectively selected by two off-chip
dense-wavelength-demultiplexer (DWDMs) filters, and following items
of the state are post-selected with a 25\% probability:

\begin{eqnarray*}
\left|\phi\right\rangle  & = & \frac{1}{\sqrt{2}}[\left|1_{s}0_{i}\right\rangle _{tt}\left|0_{s}0_{i}\right\rangle _{tb}\left|0_{s}1_{i}\right\rangle _{bt}\left|0_{s}0_{i}\right\rangle _{bb}\\
 &  & -e^{2i\theta_{SS}}\left|0_{s}0_{i}\right\rangle _{tt}\left|1_{s}0_{i}\right\rangle _{tb}\left|0_{s}0_{i}\right\rangle _{bt}\left|0_{s}1_{i}\right\rangle _{bb}]\\
\end{eqnarray*}
\begin{equation}
\end{equation}

Reform the state using the logical qubit representation as: 

\begin{equation}
\left|\phi\right\rangle =[\left|0\right\rangle _{s}\left|0\right\rangle _{i}-e^{2i\theta_{SS}}\left|1\right\rangle _{s}\left|1\right\rangle _{i}]/\sqrt{2}
\end{equation}

where $|0\rangle_{s}$ ($|0\rangle_{i}$) and $|1\rangle_{s}$ ($|1\rangle_{i}$)
denote signal (idler) photon\textquoteright s path states in two waveguides.
When $\theta_{SS}$ is chosen to be (\textit{n}+1/2)\textgreek{p} or \textit{n}\textgreek{p}
for an integer \textit{n}, we respectively obtain the two Bell states: 

\begin{equation}
\left|\Phi\right\rangle ^{\pm}=[\left|0\right\rangle _{s}\left|0\right\rangle _{i}\pm\left|1\right\rangle _{s}\left|1\right\rangle _{i}]/\sqrt{2}
\end{equation}

\begin{figure*}[t]
\begin{centering}
\includegraphics[scale=0.88]{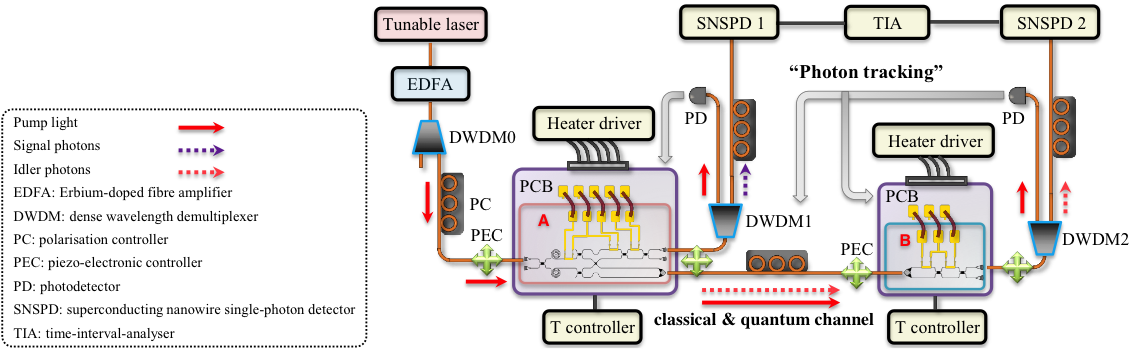}
\par\end{centering}

\centering{}\protect\caption{Experimental setup for the distribution of entanglement between integrated
Si photonic chips, chip-A and chip-B.}
\end{figure*}

\section{{Devices characterisation}}

Figure S1 shows the experimental setup for the distribution of entanglement
between two integrated silicon photonic devices. Bright light around
a wavelength of 1555.5 nm was collected from a tunable laser (Tunics-BT)
and further amplified using a high-power EDFA (Pritel). The amplified
spontaneous emission (ASE) noise was suppressed using a DWDM0 (Openti)
filter. Next, the pump light was injected into the chip-A by
using an 8-channel single-mode fibre array (OZ-Optics) with a 127
\textgreek{m}m pitch and a 10 degree polished angle. A fibre polarisation
controller (PC) was used to guarantee that TE-polarised light was
launched into the device through the 1D TE grating coupler. Entangled
photon-pairs were produced on the chip-A, and coherently distributed
to the chip-B via a 10-m single-mode fibre. Polarisation rotation
in the fibre channel was compensated by using another PC. Two off-chip
DWDMs with a 200 GHz channel space and 1 nm 1dB-bandwidth were used
to separate signal and idler photons. We selected the photons which
are equally 3-channels away from the pump, that is $\lambda_{p}-\lambda_{s}$
=$\lambda_{i}-\lambda_{p}$ = 4.8nm. This gave a high extinction ratio
to efficiently remove the remained pump from photons. Moreover, DWDM1
after the chip-A only picked up the signal photons with $\lambda_{s}$=
1550.7 nm, while DWDM2 after the chip-B only selected the idler
photons with $\lambda_{i}$= 1560.3 nm. Note DWDMs induced $\thicksim$3dB
loss for each photon. Photons were detected using two fibre-coupled
superconducting nanowire single-photon detectors (SNSPDs) mounted
in a closed cycle refrigerator. Polarisation before SNSPDs was optimized
to maximize detection efficiency up to $\thicksim$50\%. Coincidences
were recorded by using a time-interval-analyser (PicoHarp 300) in
a 450 ps integration window, which was chosen according to the jitter-time
of SNSPDs.

\subsection{Grating couplers characterisation}

Figure S2 shows the measured spectrums of the 1D and 2D grating couplers.
Peak wavelengths of both gating couplers are dependent on the angle
between fibre array and chip, and they are both around 1555.5 nm when
the relative angle is in the range of 10\textminus12 degrees. Excess
loss of 1D and 2D grating couplers is about -4.8 dB and -7.6 dB at
the peak wavelengths, and their 1dB-bandwidths are around 27 nm and
30 nm, respectively. The 1D grating couplers
consist of a periodic 315 nm silicon layer with a 630 nm pitch. The
2D grating couplers include 10 \textmu m \texttimes{} 10 \textmu m
hole arrays with a 390 nm diameter and a 605 nm pitch (Insets of Fig.S2).
Optimised angles for the two chips are slightly
different, owing to the wavelength difference of signal and idler
photons and also fabrication deviation of the devices. 
Loss of grating couplers can be further reduced by engineering the grating structure and positioning reflection mirrors under the grating {[}5, 6{]}. 
Note that other on-chip polarisation control and diversity devices can be explored for quantum linking between chips using entanglement {[}7{]}. 

\subsection{Tomography stages characterisation}

The two chips were fixed on two copper PCBs and thermal-heaters were wired-bounded
to electric pads on the PCBs. Home-made computer-interfaced heater drivers were
used to independently control all heaters on the two chips (Figure S1).
The output optical power was recorded as a function of electric power
added on heaters, from which the relationship between states and electric
power was reconstructed. A least-squares minimisation algorithm was
used to fit the \textit{O-E} (optical power and electric power) contour
and find responding powers for different states. Figure S3 shows the calibration
results of chip-A\textquoteright s and chip-B\textquoteright s projectors,
{\small{}$A(\theta_{AZ},\theta_{AY})$ and $B(\theta_{BZ},\theta_{BY})$},
by simultaneously scanning {\small{}$\theta_{AY}$} and {\small{}$\theta_{AZ}$}
or {\small{}$\theta_{BY}$} and {\small{}$\theta_{BZ}$}
phase shifters, respectively. To calibrate chip-B\textquoteright s
{\small{}$B(\theta_{BZ},\theta_{BY})$,} the input state needs to be
known in advance otherwise it is difficult to access phase offset
of the {\small{}$\theta_{BZ}$} phase shifter. We used the 1D
TE-grating coupler as an on-chip polariser to guarantee TE-polarised
state was injected. Then we kept the fibre untouched and smoothly
switched the input state to the 2D grating coupler on the same chip.
This state is an anti-diagonal state for the 2D grating coupler. Then
we determined all phase information of chip-B's {\small{}$B(\theta_{BZ},\theta_{BY})$.}
Similarly we calibrated chip-A's {\small{}$A(\theta_{AZ},\theta_{AY})$}. 

\begin{figure}[t]
\begin{centering}
\includegraphics[scale=0.75]{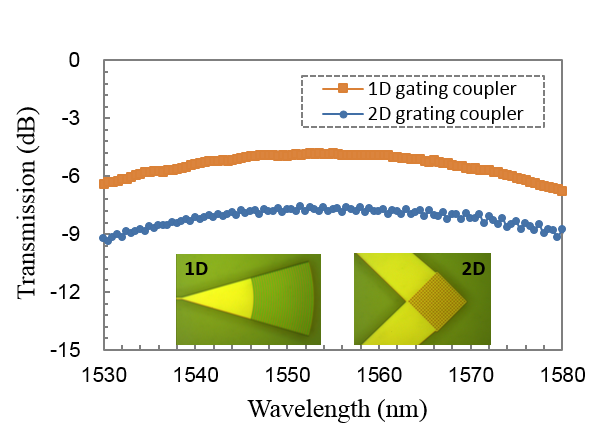}
\par\end{centering}

\centering{}\protect\caption{Spectrums of the 1D and 2D grating couplers. Peak wavelength with
maximal transmission is determined by the titled angle of fibre. These
spectrums were measured when the relative angle was optimised to be
10-12 degrees. In this case, both grating couplers work near 1555.5
nm peak wavelengths. Insets show the optical microscopy images of the grating couplers. }
\end{figure}

\begin{figure*}[t]
\begin{centering}
\includegraphics[scale=0.58]{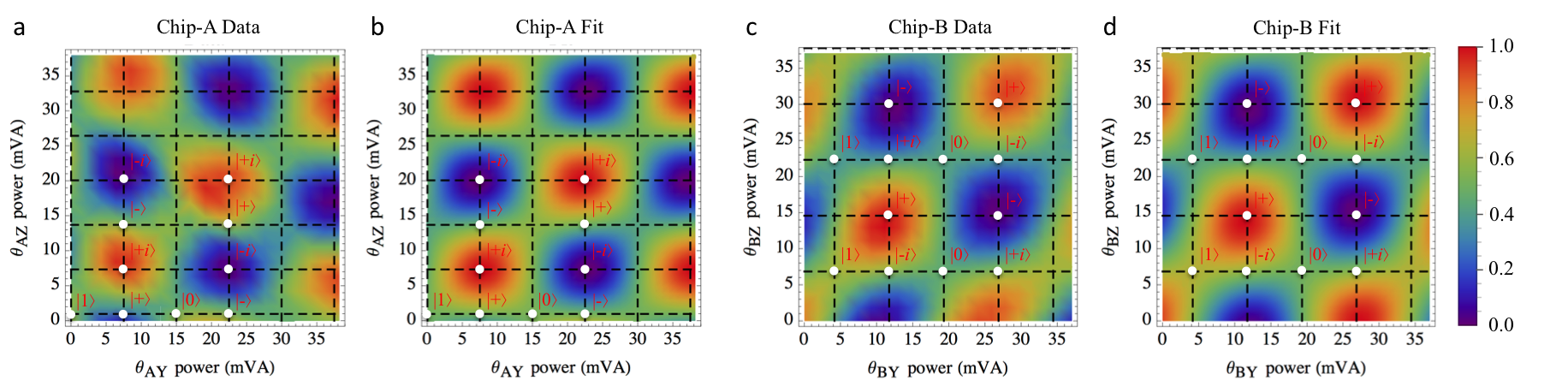}
\par\end{centering}

\centering{}\protect\caption{Calibrations of chip-A\textquoteright s and chip-B\textquoteright s
projectors. Optical power was recorded as a function of electric power
added on heaters. (a) and (b) are experimental data and fit of $A(\theta_{AZ},\theta_{AY})$
and (c) and (d) are experimental data and fit of $B(\theta_{BZ},\theta_{BY})$
in which axes are electric power added on the {\small{}$\theta_{AY}$}
and {\small{}$\theta_{AZ}$} or {\small{}$\theta_{BY}$}
and {\small{}$\theta_{BZ}$} heaters, and contour color represents
normalised optical power. This optical-electric contour was fitted
by using the least-squares minimisation algorithm to find the demanded
states. White dots at crossers denote the searched projective states
\{${|0\rangle,|1\rangle,|+\rangle,|-\rangle,|+i\rangle,|-i\rangle}$\}
in the map of electric power.}
\end{figure*}

\subsection{Systematic stability measurement}

A classical reference frame is necessary for real-life scenarios.
The phase matching condition of the SFWM processing allows the generated
photons propagate collinearly with the pump light. Then, we can use
the pump light to close a feedback loop to track single photons on
both two chips and also in the optical fibre channel, and to keep
state stability in the chip-to-chip system. The demultiplexing MMIs
on chip-A split the pump into two parts, chip-A\textquoteright s
projector and the chip-B. Half was used to feed-forwardly track
photons at the chip-A side; the other half was transmitted through
the fibre channel together with the idler photons, and further used
to track photons in the optical fibre and the chip-B side. Figure
S4 shows the stability of the chip-to-chip system. Fibre alignment
was feed-forwardly recoupled each 1 min by using a piezo electronic
stacks (PEC). It shows that path-encoded states on the two chips are
stable and polarisation-encoded states in the fibre channel are also
well-stabilised. A longer-time stability maintenance would require
an active polarisation control before the chip-A chip and also between
two chips.

\begin{figure}[t]
\begin{centering}
\includegraphics[scale=0.7]{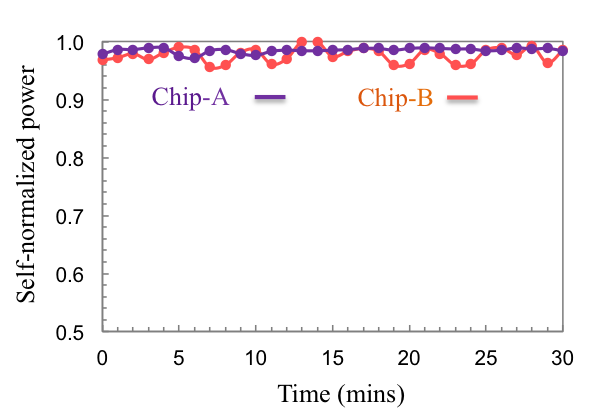}
\par\end{centering}

\centering{}\protect\caption{Stability of the chip-to-chip system. Curves represent the measured optical power and self-normalised, as a function of time. 
Purple one is obtained at chip-A\textquoteright s port D1, and pink one is measured at chip-B\textquoteright s
port D3. $A(\theta_{AZ},\theta_{AY})$ and
$B(\theta_{BZ},\theta_{BY})$ on both two chips were set as the $|+\rangle$
projectors. Fibre alignment was feed-forwardly recoupled each 1min.}
\end{figure}

\begin{figure*}[t]
\begin{centering}
\includegraphics[scale=0.65]{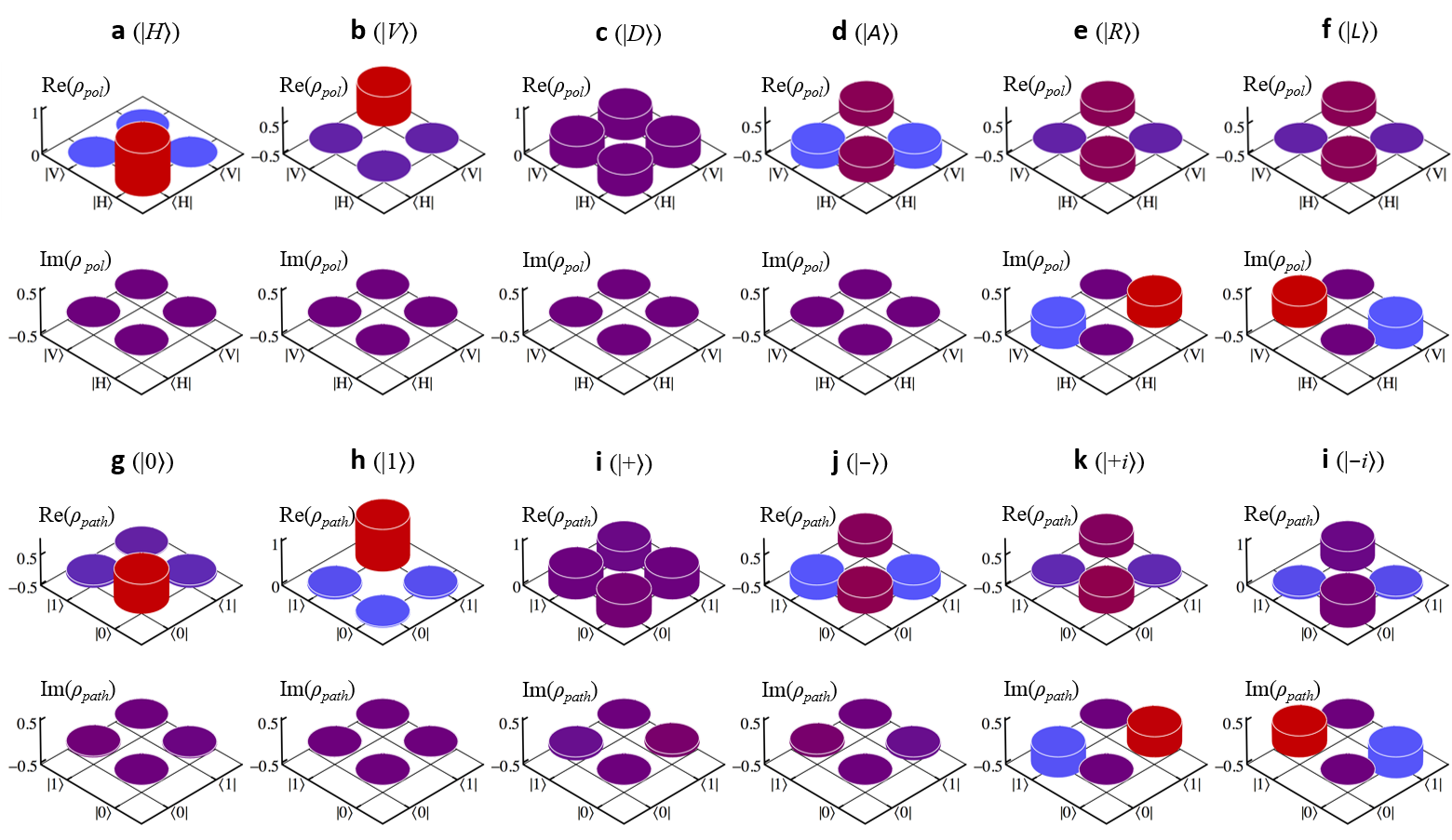}
\par\end{centering}

\centering{}\protect\caption{Density matrix of ideal polarisation-encoded states and reconstructed
path-encoded states. (a)-(f). Ideal density matrix $\rho_{pol}$ of the
polarisation-encoded states, |$H\rangle$, $|V\rangle$, $|D\rangle$,
$|A\rangle$, $|R\rangle$, and$|L\rangle$, prepared in bulky optics.
(g)-(i). Reconstructed density matrix $\rho_{path}$ of the six path-encoded
states, $|0\rangle$, $|1\rangle,$ $|+\rangle,$ $|-\rangle,$ $|+i\rangle$
and $|-i\rangle$, measured on chip. The state fidelities between
$\rho_{path}$ and $\rho_{pol}$ are estimated to be 98.66\%, 98.38\%,
99.34\%, 99.44\%, 97.66\%, and 99.46\%, respectively.}
\end{figure*}

\section{State and process tomography}

Initial polarisation encoded states, \{$|H\rangle$, $|V\rangle$,
$|D\rangle$, $|A\rangle$, $|R\rangle$, $|L\rangle$\}, were prepared
in bulk optics and injected to a copy of the chip-B. The converted
path-encoded states, \{$|0\rangle$, $|1\rangle$, $|+\rangle$, $|-\rangle$,
$|+i\rangle$, $|-i\rangle$\}, were analysed on-chip by implementing
state tomography {[}8{]}. Each state was measured in six basis states
\{$|0\rangle$, $|1\rangle$, $|+\rangle$, $|-\rangle$, $|+i$,
$|-i\rangle$\}. A maximum likelihood search algorithm was used to
reconstruct the most likely legitimate state from measurement results.
Figure S5 shows the ideal polarisation-encoded states $\rho_{pol}$
and the six reconstructed path-encoded states $\rho_{path}$, and
the state fidelities defined as the overlapping between $\rho_{pol}$
and $\rho_{path}$. By subjecting polarisation-encoded states into
the PPC and measuring output path-encoded states (determined using
state tomography), we determinate the process matrix \textgreek{q}
of the PPC processing for a fixed set of operators $E_{i}$, where
$E_{1}$, $E_{2}$, and $E_{3}$ are simply chosen to be the Pauli
operators and $E_{0}$ to be the Identity operator. Twelve parameters
in the PPC\textquoteright s process matrix \textgreek{q} need to be
determined. Four ideal polarisation states $\rho_{pol}$, for example,
\{$|H\rangle$, $|V\rangle$, $|D\rangle$, $|R\rangle$\}, were chosen
as the input, and reconstructed path-encoded states $\rho_{path}$
\{$|0\rangle$, $|1\rangle$, $|+\rangle$, $|+i\rangle$\}, were
used as the output. The process matrix \textgreek{q} is described
explicitly as {[}8{]}:

\begin{equation}
\chi=\Lambda\Omega\Lambda
\end{equation}
where the matrix \textgreek{W} is determined by the reconstructed
density matrix $\rho_{path}$ of ${|0\rangle,|1\rangle,|+\rangle,|+i\rangle}$
and the matrix \textgreek{L} is described as: 

\begin{equation}
\chi=\left[\begin{array}{cc}
I & X\\
X & -I
\end{array}\right]\frac{1}{2}
\end{equation}

\section{Entanglement correlation coefficients}

Correlation coefficients $\left\langle A_{i},B_{i}\right\rangle $
or $\left\langle A_{i}(\theta_{AY}),B_{i}(\theta_{BY})\right\rangle $
inside the Bell-CHSH inequality are defined as the normalised correlation
value of the measured coincidences data, shown as below {[}9, 10{]},
equation (D1). In our experiment, we respectively measured the coincidences
with a chip-A\textquoteright s $\theta_{AY}$ set of \{0, \textgreek{p}/2,
\textgreek{p}, 3\textgreek{p}/2\} and a chip-B\textquoteright s $\theta_{BY}$
set of \{\textgreek{p}/4, 3\textgreek{p}/4, 5\textgreek{p}/4, 7\textgreek{p}/4\},
to get the correlation coefficients of $\left\langle A(0),B(\pi/4)\right\rangle $,
$\left\langle A(0),B(\pi3/4)\right\rangle $, $\left\langle A(\pi/2),B(\pi/4)\right\rangle $,
and $\left\langle A(\pi/2),B(\pi3/4)\right\rangle $. Figure S6 shows
the measured correlation coefficients of the two Bell states. These
correlation coefficients were consequently used to calculate the Bell-CHSH
\textit{S} parameter using equation (D2). 

\begin{figure*}[t]
\begin{centering}
\includegraphics[scale=0.72]{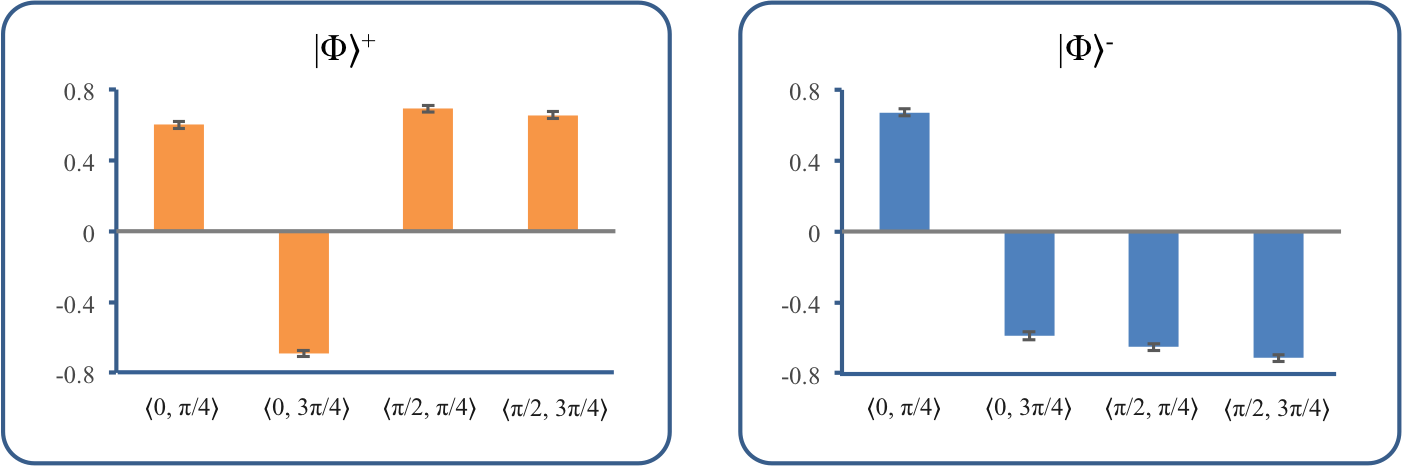}
\par\end{centering}

\centering{}\protect\caption{Measured correction coefficients of two Bell states $|\Phi\rangle^{+}$
and $|\Phi\rangle^{-}$ after distributed across the chip-A and chip-B
chips. Coincidences of each measurement were accumulated for 60s.
Accidental coincidences are subtracted for all data. Standard deviation
of the correction coefficients are calculated from an evolution of
Poissonian photon statistics. }
\end{figure*}

\begin{widetext} 
\begin{eqnarray} 
\ \left\langle A(\theta_{AY}),B(\theta_{BY})\right\rangle= 
\frac{[C(\theta_{AY},\theta_{BY})+C(\theta_{AY}+\pi,\theta_{BY}+\pi)-C(\theta_{AY},\theta_{BY}+\pi)-C(\theta_{AY}+\pi,\theta_{BY})]}{[C(\theta_{AY},\theta_{BY})+C(\theta_{AY}+\pi,\theta_{BY}+\pi)+C(\theta_{AY},\theta_{BY}+\pi)+C(\theta_{AY}+\pi,\theta_{BY})]} \\ 
S=\parallel\left\langle A_{1}(\theta_{AY}),B_{1}(\theta_{BY})\right\rangle +\left\langle A_{1}(\theta_{AY}),B_{2}(\theta_{BY})\right\rangle +\left\langle A_{2}(\theta_{AY}),B_{1}(\theta_{BY})\right\rangle -\left\langle A_{2}(\theta_{AY}),B_{2}(\theta_{BY})\right\rangle \parallel
\
\end{eqnarray} 
\end{widetext}

\section*{{\large{}References}}

{\small{}1. Clemmen, S. et al. Continuous wave photon pair generation
in silicon-on-insulator waveguides and ring resonators. Opt. Express,
17, 16558\textendash 16570 (2009). }{\small \par}

{\small{}2. Sharping, J. E. et al. Generation of correlated photons
in nanoscale silicon waveguides. Opt. Express, 14, 12388\textendash{}
12393 (2006). }{\small \par}

{\small{}3. Silverstone, J. W. et al. On-chip quantum interference
between silicon photon-pair sources. Nat. Photon. 8, 104\textendash 108
(2014). }{\small \par}

{\small{}4. Silverstone, J. W. et al. Qubit entanglement on a silicon
photonic chip. arXiv:1410.8332 {[}quant-ph{]} (2014). }{\small \par}

{\small{}5. Vermeulen, D. et al. High-efficiency fiber-to-chip grating
couplers realized using an advanced CMOS-compatible Silicon-On-Insulator platform. Opt. Express.18. 18278\textendash18283 (2010)}{\small \par}

{\small{}6. Ding, Y. et al. Fully etched apodized grating coupler on the SOI platform with -0.58 dB coupling efficiency. Opt. Lett. 39. 5348\textendash 5350 (2014)}{\small \par}

{\small{}7. Dai, D. et al. Passive technologies for future large-scale photonic integrated circuits on silicon: polarization handling, light non-reciprocity and loss reduction. Light: Science \& Applications. 1 (2012)}{\small \par}

{\small{}8. Nielsen, M. \& Chuang, I. Quantum computation and quantum
information. (Cambridge Univ. Press, 2000). }{\small \par}

{\small{}9. Clauser, J. F. et al. Proposed experiment to test local
hidden-variable theories. Phys. Rev. Lett. 23, 880\textendash 884
(1969). }{\small \par}

{\small{}10. Aspect, A. et al. Experimental Tests of Realistic Local
Theories via Bell's Theorem, Phys. Rev. Lett. 47, 460\textendash 463
(1981).}{\small \par}
\end{document}